%% file: acl_latex.tex
\pdfoutput=1

\documentclass[11pt]{article}

\usepackage[final]{acl}
\usepackage{graphicx}
\usepackage{arydshln}
\usepackage{times}
\usepackage{amsmath} 
\usepackage{amsfonts} 
\usepackage{amssymb}
\usepackage{graphicx}
\usepackage{subcaption}  
\usepackage{url}
\usepackage{latexsym}

\usepackage[T1]{fontenc}

\usepackage[utf8]{inputenc}

\usepackage{microtype}

\usepackage{inconsolata}
\usepackage{enumitem}
\usepackage{graphicx}
\usepackage{colortbl}
\usepackage{multirow}
\usepackage{xcolor}
\usepackage{tcolorbox} 
\usepackage{booktabs}
\usepackage{ulem}

\newcommand{\bk}[1]{{\color{black}{#1}}}
\newcommand{\bkq}[1]{{\color{black}{#1}}}
\newcommand{\zjz}[1]{{\color{black}{#1}}}

\title{Decoding Matters: Addressing Amplification Bias and Homogeneity Issue for LLM-based Recommendation}

\author{
Keqin Bao$^{1,2}$\thanks{\hspace{2mm}Work is done during internship at Huawei Inc.} \:\:\: \textbf{Jizhi Zhang}$^{1}$ \:\:\: \textbf{Yang Zhang}$^{3}$\thanks{\hspace{2mm}Corresponding Author} \:\:\:\\ \textbf{Xinyue Huo}$^{1}$ \:\:\: \textbf{Chong Chen}$^{3}$ \:\:\:\textbf{Fuli Feng}$^{1\dagger}$ \:\:\:\\
$^{1}$University of Science and Technology of China, \\
$^{2}$Huawei Inc. \:
$^{3}$National University of Singapore \\
\texttt{\{baokq, cdzhangjizhi, xinyueh\}@mail.ustc.edu.cn}, \\
\texttt{chenchong55@huawei.com},
\texttt{\{zyang1580, fulifeng93\}@gmail.com} 
}

\begin{document}
\maketitle
\input{latex/0abstract}
\input{latex/1intro}

\input{latex/2related_work}
\input{latex/3preliminary}
\input{latex/4method}
\input{latex/5experiment}

\input{latex/6analysis}
\input{latex/7conclusion}

\section*{Acknowledgments}
This work is supported by the National Natural
Science Foundation of China (62272437), and the advanced computing resources provided by the Supercomputing Center of the USTC.
\section*{Limitations}
This paper primarily focuses on the issues and challenges that arise when deploying LLMs in the domain of recommendation systems, particularly at the decoding phase, and offers a viable and effective solution. However, our study has several limitations:
1) The experiments conducted in this study are solely based on the Qwen1.5-1.8B model.
2) The current investigation emphasizes performance enhancement in sequential recommendation tasks. Although the method introduced in this paper does not significantly increase the inference time compared to existing \zjz{recLLM} methods, it encounters efficiency issues in inference when applied to real-world recommendation scenarios.
A potential solution to address the inference efficiency problem is to eliminate the generation of ghost tokens. By integrating existing generation technologies like vLLM~\cite{kwon2023efficient}, it is possible to significantly reduce the number of inference tokens and, consequently, the inference time. However, due to the limitations of our experimental setup and the fact that our primary focus was not on this topic, it was not experimentally tested in this paper.

\section*{Ethical Considerations}
In this paper, we introduce $D^3$ to mitigate amplification bias and the problem of homogeneity in existing LLM-based recommendation decoding strategies, without introducing new ethical dilemmas. We employ publicly available data while carefully avoiding sensitive information. Nevertheless, as recommendations depend on user behavior data, privacy issues may arise. These concerns can be managed by securing user consent. Moreover, the usage of LLMs might perpetuate unseen societal biases. We recommend comprehensive risk evaluations and caution users about the potential risks associated with model deployment.

\bibliography{anthology, custom}
\appendix

\input{latex/8appendix}

\end{document}

%% file: latex/0abstract.tex
\begin{abstract}
Adapting Large Language Models (LLMs) for recommendation requires careful consideration of the decoding process, given the inherent differences between generating items and natural language.
Existing approaches often directly apply LLMs' original decoding methods. However, we find these methods encounter significant challenges: 1) amplification bias—where standard length normalization inflates scores for items containing tokens with generation probabilities close to 1 (termed ghost tokens), and 2) homogeneity issue—generating multiple similar or repetitive items for a user.
To tackle these challenges, we introduce a new decoding approach named \textit{Debiasing-Diversifying Decoding} (\textit{$D^3$}). \textit{$D^3$} disables length normalization for ghost tokens to alleviate amplification bias, and it incorporates a text-free assistant model to encourage tokens less frequently generated by LLMs for counteracting recommendation homogeneity. 
Extensive experiments on real-world datasets demonstrate the method's effectiveness in enhancing accuracy and diversity. The code is available at \url{https://github.com/SAI990323/DecodingMatters}.


\end{abstract}

%% file: latex/1intro.tex
\section{Introduction}

Researchers have endeavored to adapt Large Language Models (LLMs) for recommender systems, seeking to harness the remarkable abilities of LLMs to improve recommendation performance~\cite{TALLRec, recommendation_instruction, wei2023llmrec, harte2023leveraging}. Numerous adaptation solutions have been proposed, with some showcasing notable success. Among these, using LLMs to perform recommendations in a generative manner is particularly promising~\cite{bao2023bi, tan2024towards, lcrec}. This approach involves teaching LLMs to directly generate suitable item representations (\textit{e.g.}, item titles) as recommendations based on given instructions. By doing so, this approach aligns well with the generative nature of LLMs, potentially enabling a more effective utilization of their capabilities for recommendation.



To generate an item, LLMs must produce a sequence of tokens representing the item, necessitating multi-step decoding during generation. Existing approaches typically adopt original decoding strategies of the utilized language models, such as beam search~\cite{bao2023bi, lcrec, tan2024towards}. 
However, there are distinct disparities between generating recommended items and natural language~\cite{bao2023bi, lcrec}. For instance, in terms of the generation space, items only correspond to a specific, non-uniformly sampled subset of the entire language space; in terms of \bkq{generation results}, a set of items should be recommended, as opposed to 
only \bkq{ond appriopriate} 
language output required in typical NLP tasks. These distinctions potentially introduce new challenges for directly applying the heuristic decoding strategies.


This work delves into studying the decoding process of LLMs for generating recommendations. We identified two potential critical issues for applying the original decoding strategies of LLMs:
\begin{itemize}[topsep=0pt,itemsep=2pt,parsep=0pt,leftmargin=*]
    \item 
    Amplification Bias: Due to the non-uniform distribution of items in the language space, some items may contain tokens with generation probabilities close to 1 under certain conditions (referred to as \bkq{ghost} tokens). Existing decoding methods tend to enhance the scores for these items. Typically, these methods utilize length normalization to counteract the length bias during generation~\cite{wu2016google, vaswani2017attention} --- longer sequences tend to have lower probabilities than shorter sequences merely due to the multiplication of probabilities for each token. However, when \bkq{ghost} tokens appear, multiplying their generation probabilities (near 1) doesn't significantly decrease the final score, but length normalization is still applied, resulting in score amplification.

    \item 
    Homogeneity Issue: With the original decoding methods, LLMs often produce items with similar structures and content when offering multiple recommendations to a user, as textually similar sequences typically receive similar scores. For example, when suggesting products, the model may recommend several items from the same series or category (e.g., ``PlayStation 3'' and ``PlayStation 4''). Moreover, the model frequently repeats item features based on past user interactions, due to inheriting the match-and-copy mechanism of LLMs~\cite{olsson2022context,wang2022interpretability}.
\end{itemize}

Both amplification bias and homogeneity issues would undeniably exert substantial influences on recommendation quality, impacting factors such as accuracy and diversity. 
Addressing these issues is crucial. 
However, this endeavor is undoubtedly non-trivial. 
As discussed, the emergence of amplification bias is intricately linked with the process of addressing the length bias of LLMs' generation. 
\bkq{Therefore, it's crucial to navigate addressing amplification bias without reintroducing the length bias. }
Meanwhile, to address the homogeneity issue, we must enhance the generation probability of tokens in LLM that receive less attention. 
However, there is a lack of clear guidance on how to balance the original predictions with enhancing less attention tokens and which tokens to select.

\bkq{
To this end, we introduce a novel strategy called \textit{Debiasing-Diversifying Decoding} ($D^3$). To mitigate amplification bias while still addressing length bias, $D^3$ considers selectively applying length normalization to tokens while excluding it for ghost tokens. When implementing, we find that excluding ghost tokens results in relatively uniform item lengths, making the remaining length normalization unnecessary. Consequently, \textit{$D^3$} removes all length normalization during decoding, facilitating its implementation. To tackle the homogeneity issue, \textit{$D^3$} incorporates scores from a text-free assistant model at each decoding step to guide token generation. This helps avoid generating similar and repetitive items, as the text-free model is not influenced by repetitive text and can provide meaningful non-repeated/non-similar token suggestions based on its recommendation capabilities.
}


In total, our contribution is as follows:
\begin{itemize}[topsep=0pt,itemsep=0pt,parsep=0pt,leftmargin=*]
    \item 
    We highlight the importance of decoding strategies for LLMs' generation in recommendation scenarios and identify two critical challenges: amplification bias towards items with ghost tokens and the homogenous issues of generating similar and repeated items.
    \item We point out the ghost token and the repetition phenomenon and propose to mitigate this issue by removing the length penalty and utilizing a text-free assistant model.
    \item Extensive experiments have demonstrated that our methods can simultaneously enhance the accuracy and diversity of recommendations.
\end{itemize}

%% file: latex/2related_work.tex
\section{Related Work}
In this section, we will briefly introduce the development and status of the related fields of this article, mainly including LLMs for recommender systems and decoding strategies for language models.

\subsection{Large Language Models for Recommender Systems} 
The remarkable capabilities demonstrated by LLMs have sparked interest within the recommender systems community, leading to a surge in research aimed at integrating LLMs into these systems \cite{llmrecsurvey1, llmrecsurvey2, tutorial1}. 
This research can be broadly divided into two streams.
The first stream involves directly utilizing LLMs as encoders to represent items, which are then input into conventional recommendation models \cite{encoder1, encoder2}. 
However, the prevalent LLMs are primarily decoder-only architectures, optimized for next-token-prediction, which is not conducive to information encoding. 
This limitation potentially constrains the full potential of LLMs in recommendation tasks.

The second stream of research focuses on harnessing the generative power of LLMs to produce recommendations directly~\cite{lin2023multi, liao2023llara, recmind, zhang2023collm, zhang2024prospect, zhang2024text}. 
Despite this approach, it has been observed that LLMs have limited exposure to recommendation-specific data during pre-training, necessitating fine-tuning for effective application in this domain. 
While methods of this nature are on the rise, they often prioritize improving LLMs' recommendation performance through training enhancements, overlooking a crucial aspect: the detailed examination of the models' recommendation outputs and the considerations for the decoding phase. 


\subsection{Decoding in Language Model} 
In the domain of open-ended text generation, the issue of diversity has garnered significant attention, primarily due to the inherent dependency of language models on maximizing probability during the generation process~\cite{welleck2019neural,holtzman2019curious}. 
To address this, a common approach is to adjust the temperature parameter. 
This method modifies the sharpness of the probability distribution of tokens at each step of generation, reducing the likelihood of exceedingly similar outcomes due to overly high probabilities of specific tokens being chosen. 
Beyond this, Diverse Beam Search (DBS)~\cite{vijayakumar2016diverse} introduces the concept of beam groups, partitioning the generation process to manage similarity across different groups, thereby increasing the diversity of the generated results.
Currently, due to the differences in the decoding space, using LLMs for recommendations presents unique challenges. There is a lack of comprehensive discussion on the decoding stage in this field. In this paper, we conduct the first exploratory analysis and optimization of the decoding process and the recommendations.


%

The second stream of research focuses on harnessing the generative power of LLMs to produce recommendations directly~\cite{lin2023multi, liao2023llara, recmind, zhang2023collm, zhang2024text}. 
Despite this approach, it has been observed that LLMs have limited exposure to recommendation-specific data during pre-training, necessitating fine-tuning for effective application in this domain. 
While methods of this nature are on the rise, they often prioritize improving LLMs' recommendation performance through training enhancements, overlooking a crucial aspect: the detailed examination of the models' recommendation outputs and the considerations for the decoding phase. 

%% file: latex/3preliminary.tex
\section{Preliminary}
In this section, we will introduce the background knowledge of the decoding process of current LLM-based recommender (recLLM) methods.

Considering the decoding process, the model takes an instruction input and a sequence of user historical interactions to generate a suitable token sequence representing a recommended item. Our input, of length $n$, is denoted as $x = x_1 \cdots x_n$, where each $x_i$ is a token from the model's vocabulary $V$. During decoding, an item is represented by generating a sequence of tokens having length $m$, denoted as $y = y_{1} \cdots y_{m}$. 
At decoding time, tokens are generated iteratively, one at a time, conditioned on the previous context and expect to select the output with the highest probability:
\begin{align}
p(y|x) = \prod_{i=1}^{m} p(y_i | x, y_{<i}),
\end{align}
where $p(y_i | x, y_{<i})$ represents the probability distribution of the next token. 
However, given the impracticality of traversing all possible scenarios to calculate this probability, the greedy search can apply to select the token $x_i$ with the highest probability at each step.

\paragraph{\textbf{Beam Search.}}
Due to the limitations of greedy decoding methods, which can only generate a single item and often fail to find the optimal sequence of items~\cite{holtzman2019curious}, researchers currently using LLMs for recommendation typically employ beam search methods to generate multiple items simultaneously~\cite{lcrec, tan2024towards}.
In detail, the formula for calculating the score for a hypothesis $h$ at step $ t $ in beam search is: 
\begin{align}
\mathcal{S}(h_{\leq t}) = \mathcal{S}(h_{\leq t-1}) + \log(p(h_t|x, h_{\leq t-1})),
\label{eq:2}
\end{align}
where $\mathcal{S}(h_{\leq t})$ is the score of the hypothesis up to step $t$ used for selecting hypothesis, $p(h_t|x, h_{\leq t-1})$ is the conditional probability of the token $h_t$ given the previous hypothesis $ h_{\leq t-1} $ and the input $x$.

The beam search keeps track of the top $B$ hypotheses at each step based on their scores, where $B$ is the beam width. It then expands each of these hypotheses by considering the $k$ most likely tokens. The new hypotheses are scored according to the above formula, and the top $B$ hypotheses are retained for the next step. This process continues until an end-of-sequence token is predicted or the maximum sequence length is reached. Moreover, in natural language generation, to avoid repetitiveness and overly long outputs, a length normalization term is typically added after the generation is complete.
\begin{align}
\mathcal{S}(h) = \mathcal{S}(h) / h_L^{\alpha},
\label{eq:3}
\end{align}
where $h_L$ is the length of the $h$, $\alpha$ is the hyper-parameters to control the length penalty. Once the beam search is completed, the top $B$ hypotheses are selected as the final output $[y_1, y_2, ..., y_B]$.

%% file: latex/4method.tex
\section{Issues for Existing Decoding Method}
\input{latex/figure/figure_analysis}

In this section, we elaborate on the amplification bias and homogeneity issues of the existing decoding method, clarifying our motivation.


\subsection{Amplification Bias} 
As discussed, due to the non-uniformity of the item generation space, there are ghost tokens whose generation becomes deterministic once their preceding tokens have appeared. The model easily learns this deterministic characteristic during training and subsequently assigns a near 1-generation probability to these tokens during the generation phase. For Eq. (\ref{eq:2}) and Eq. (\ref{eq:3}), it is evident that these tokens minimally impact the value of $\mathcal{S}$ but increase the length $h_L$. This means they act like ghosts, occupying a sequence position without changing the score $\mathcal{S}$. Consequently, if the length normalization in Eq. (\ref{eq:3}) is still applied, the final score would be improperly amplified.

\subsection{Homogeneity Issue}
Next, we illustrate the homogeneity issue in generated results by analyzing the recommendation similarity within a recommended list and their relative repetition compared to historical data.

\paragraph{Recommendation Similarity.}
We note that the top-10 recommendations from recLLM often begin with similar tokens, reducing diversity. To compare, we examine the first 5 tokens in these recommendations against those from conventional models\footnote{For traditional models, we replaced the item IDs they produced with their corresponding textual representations.}.
\bkq{Furthermore, we also assess the category diversity of the recommended items.}
As depicted in Figure~\ref{fig:sub1},  we utilized BLEU to measure the textual similarity among recommendation results, and Entropy to assess the category diversity. 
\bkq{It is evident that, compared to traditional models, the results from recLLMs exhibit higher degrees of similarities. }
This suggests that, despite their impressive performance, LLMs tend to generate multiple homogeneous recommendations, given a user. 
This homogeneity might be attributed to the excessive scores given to the first few similar tokens by the beam search, which makes it challenging for more diverse outcomes to be included in the final candidate generations~\cite{vijayakumar2016diverse}.

\paragraph{Repetition Phenomenon.} 

Looking deeper, we observe that certain tokens repetitively appear both in the recommended items and historically interacted items. Thus, we continue our analysis to determine the extent of similarity between the recommended items and the user’s historical interactions.
Similarly, we computed the textual and categorical similarities between the recommendations generated by LLMs/traditional models and users' historical interaction records.
As illustrated in Figure~\ref{fig:sub3}, we observe that similar to the preceding findings, the recommendations based on recLLM also exhibit higher similarity to users' historical interaction sequences. This suggests that while incorporating textual information offers certain benefits, it also introduces a new bias: the model tends to copy from the preceding text~\cite{olsson2022context}, leading to a prevalence of similar items in the recommendations.

\section{Debiasing-Divesifying Decoding}
Next, we introduce Debiasing-Diversifying Decoding, which incorporates two new designs into the existing method to address the two issues.


\paragraph{Remove Amplification Bias.} 
To address amplification bias, according to its source, an intuitive approach is to exclude ghost tokens when calculating the sequence length for normalization in Eq.~\eqref{eq:3}. In essence, apply length normalization only to normal tokens.
However, our analysis reveals that removing these tokens results in item token sequences with a much uniform length distribution, making length normalization for the remaining tokens unnecessary\footnote{The detailed analysis can be seen in Appendix \S\ref{appendix:sec3}}.  
Therefore, we opt to directly eliminate the length normalization to neutralize the impact of the ghost token and remove the amplification bias for the decoding.


\paragraph{Address Homogeneity Issue.}

In light of the Homogeneity analysis, we conclude that during beam search, numerous potential recommendation outcomes are prematurely pruned due to the dominance of certain tokens with exceptionally high scores~\cite{vijayakumar2016diverse}. Consequently, this not only impairs the recommendation quality but also the recommendation diversity.
To address this issue, it is imperative to refine the model’s scoring mechanism to ensure a more soft distribution of scores, thereby broadening the range of choices available during the generation phase. Hence,  we aim to adjust the score of each candidate token at each step to help recLLM better generate the item to be recommended.


To achieve this objective, the key is to boost the scores of tokens that are meaningful but underestimated by recLLM, while avoiding excessive disregard of tokens with higher scores from recLLM to maintain recommendation performance. 
Relying solely on recLLM's predictions to accomplish this is challenging. 
Therefore, we propose leveraging an additional text-free model to assist. Although this model has poor recommendation abilities, it can still provide meaningful recommendation proposals, which are less prone to textual similarity issues, for recLLM reference.

In detail, we leverage the text-free model to generate additional scores for tokens at each decoding step, using these scores to refine the token scores of recLLM. Let $\mathcal{I}$ denote the set of all items, and $\mathcal{I}_i$ denote the $i$-th item.
We denote the probability of the text-free model to recommending $\mathcal{I}_i$ as $p_{TF}(\mathcal{I}_i)$. At a decoding step, given the hypothesis $h_{<t}$ that has already been produced, we need to generate a new token $h_t$ based on it. For which, only a subset of items is eligible for the generation, that is, the items matching with $h_{<t}$, denoted as $\mathcal{I}_{h{<t}}$. Based on this set, we define the token score $\mathcal{S}_{TF}(h_{\leq t})$ of the text-free model for a potential token $h_t$ as follows:
\vspace{-1em}
\[
\scalebox{0.8}{%
\begin{minipage}{0.6\textwidth} 
\begin{align}
    \mathcal{L}_{TF}(h_t | h_{\leq t - 1}) &= log\left(\frac{\sum_{i \in \mathcal{I}_{h_{<t}, h_t}}p_{TF}(\mathcal{I}_i)}{\sum_{i \in \mathcal{I}_{h_{<t}}}p_{TF}(\mathcal{I}_i)}\right), \\
    \mathcal{S}_{TF}(h_{\leq t}) &= \mathcal{S}_{TF}(h_{\leq t  - 1}) +  \mathcal{L}_{TF}(h_t | h_{\leq t - 1}), \\
        p_{TF}(\mathcal{I}_i) &= 0 (I_i \notin \mathcal{G})
\end{align}
\notag
\end{minipage}%
}
\]
where $\mathcal{I}_{h_{<t}, h_t}$ denote the items in $\mathcal{I}_{h_{<t}}$ with $h_t$ coming after $h_{<t}$, $\mathcal{L}_{TF}(h_t | h_{\leq t - 1})$ denotes the log probability for the token by the text-free assistant model, similar to Eq.~\eqref{eq:2}.
Then the overall score function we used to guide the recLLM to generate step by step is:
\vspace{-0.5em}
\[
\scalebox{0.9}{%
\begin{minipage}{0.5\textwidth} 
\begin{align}
    \Tilde{\mathcal{S}}(h_{\leq t}) = \alpha * \mathcal{S}(h_{\leq t}) + (1 - \alpha) * \mathcal{S}_{TF}(h_{\leq t}),
\end{align}
\notag
\end{minipage}%
}
\]
where $\alpha$ is a hyper-parameter to control the assistance level of the text-free model. Notably, $ \mathcal{S}(h_{\leq t})$ is computed without the length normalization.

This formulation indicates that during each step of generation, we do not completely depend on the knowledge encoded within the recLLM. Instead, we harness logits from a text-free model, which is detached from linguistic context, to guide the recLLM in generating. By infusing text-free model inferences at every stage, we mitigate the homogeneity and redundancy that stem from the model's overdependence on language-based attributes.

%% file: latex/figure/figure_analysis.tex
\begin{figure*}[htbp]
    \centering
    \begin{subfigure}[b]{0.24\textwidth}
        \centering
        \includegraphics[width=\textwidth]{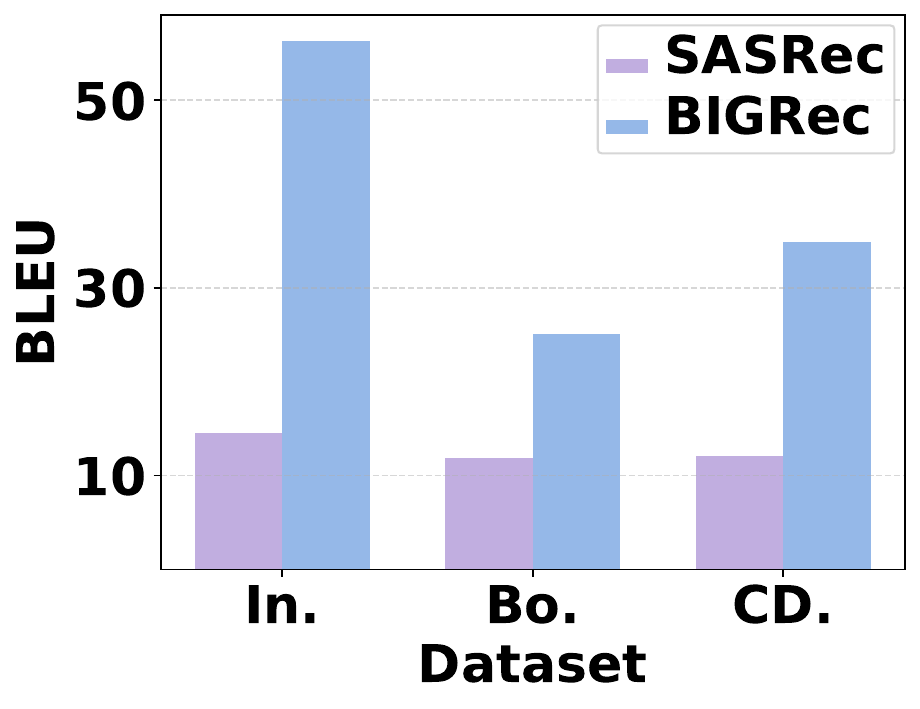}
        \vspace{-1.5em}
        \caption{}
        
        \label{fig:sub1}
    \end{subfigure}
    \hfill
    \begin{subfigure}[b]{0.23\textwidth}
        \centering
        \includegraphics[width=\textwidth]{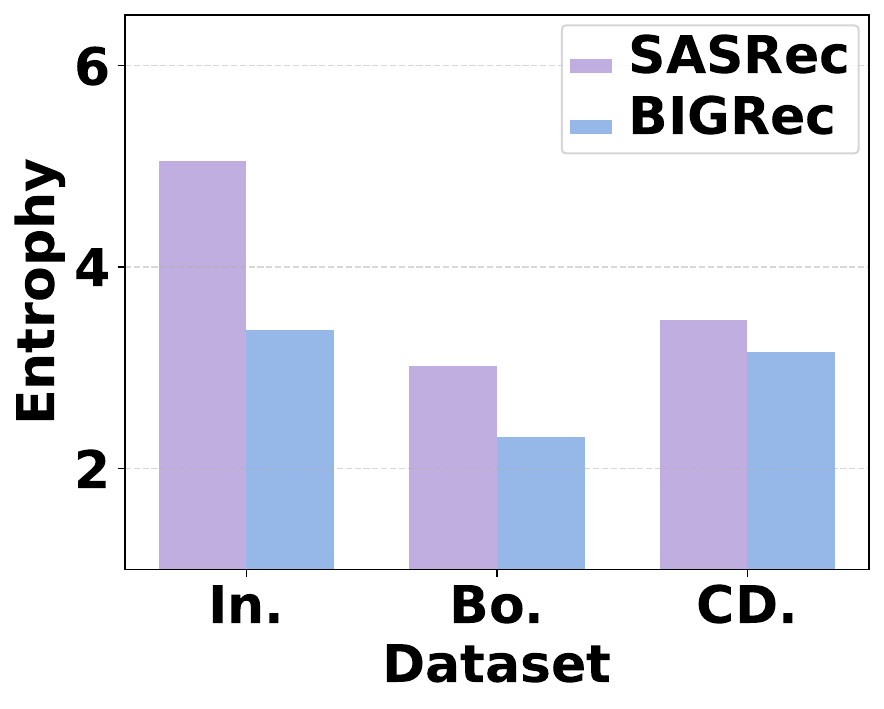}
        \vspace{-1.5em}
        \caption{}
        \label{fig:sub2}
    \end{subfigure}
    \hfill
    \begin{subfigure}[b]{0.24\textwidth}
        \centering
        \includegraphics[width=\textwidth]{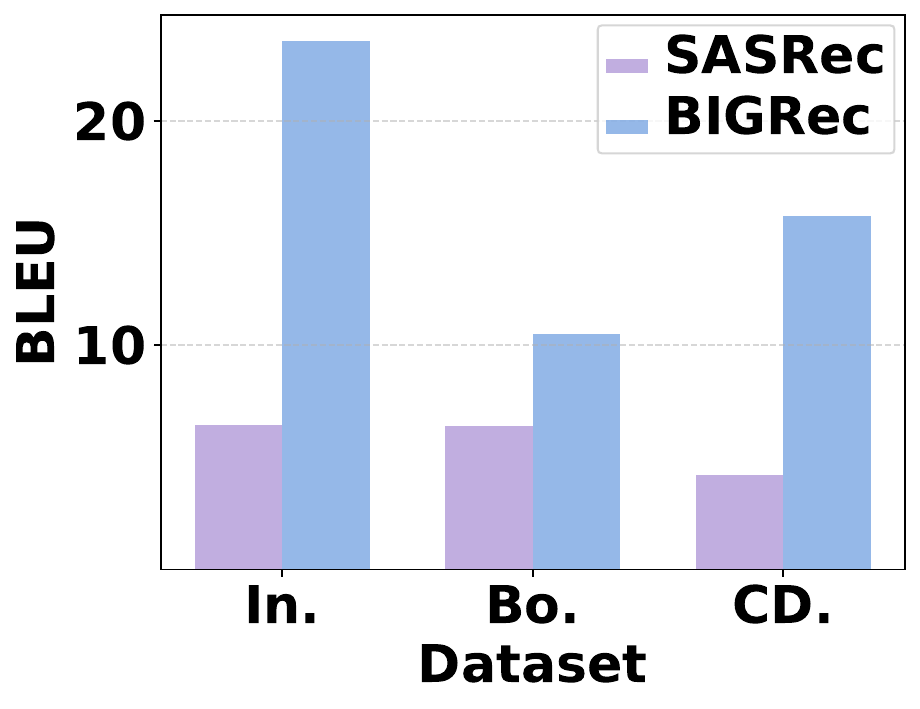}
        \vspace{-1.5em}
        \caption{}
        \label{fig:sub3}
    \end{subfigure}
    \hfill
    \begin{subfigure}[b]{0.24\textwidth}
        \centering
        \includegraphics[width=\textwidth]{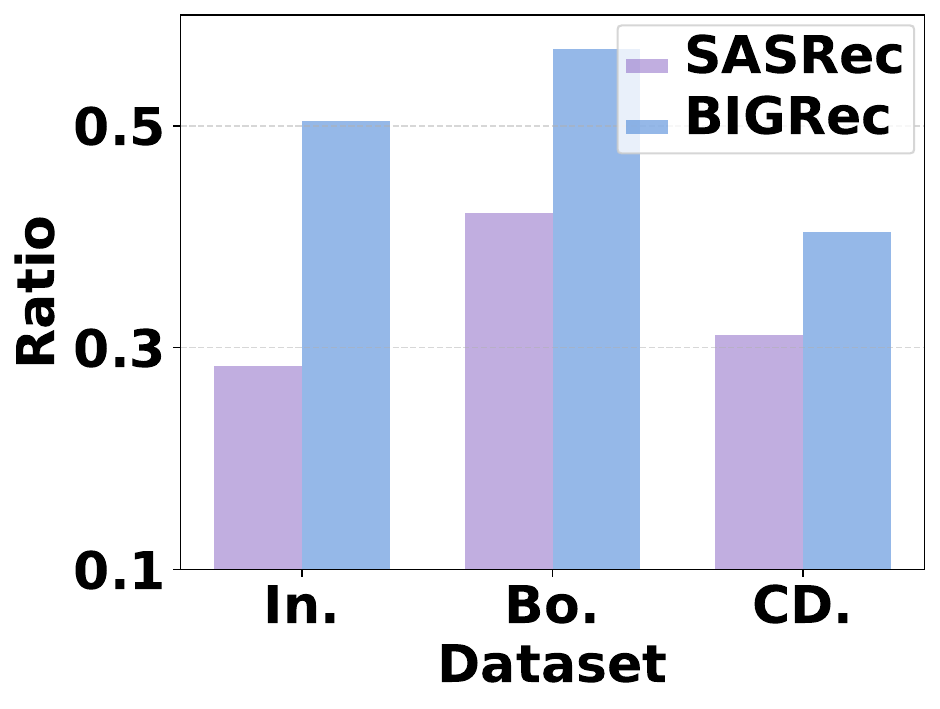}
        \vspace{-1.5em}
        \caption{}
        \label{fig:sub4}
    \end{subfigure}
    \caption{
    Homogeneity comparison of recommendation results between recLLM method BIGRec and traditional method SASRec on three datasets: \textit{Instruments} (\textit{In.}), \textit{Books} (\textit{Bo.}), and \textit{CDs} (\textit{CD.}). (a) and (b) show text similarity and category diversity (measured by entropy) for the first 5 tokens within the top 10 recommendations, where higher similarity and lower entropy indicate greater homogeneity. (c) and (d) display text similarity and category repetition in top-10 recommendations compared to historical interactions. 
    }
    \label{fig:main}
\end{figure*}

%% file: latex/5experiment.tex
\section{Experiment}
\input{latex/table/table1}
\subsection{Experimental Settings}
\subsubsection{Dataset} We conduct experiments on six real-world datasets from Amazon review data\footnote{\url{https://cseweb.ucsd.edu/~jmcauley/datasets.html\#amazon_reviews}.}, including \textit{Instruments}, \textit{CDs}, \textit{Games}, \textit{Toys}, \textit{Sports}, and \textit{Books}. All datasets contain user review data from May 1996 to October 2018. 
To preprocess the data, we follow the strategy outlined in the BIGRec paper to truncate the dataset based on time information, considering the high cost of training LLMS. Moreover, we filter out unpopular users and items with fewer than five interactions and set the maximum item sequence length to 10 to meet the baseline requirements. Details of the preprocessing steps and statistics of the processed datasets can be found in Appendix~\S\ref{appendix:A}.

\subsubsection{Evaluation Protocal}
To evaluate the model's top-K recommendation performance (accuracy), we use two commonly used metrics: Hit Ratio (HR@K) and Normalized Discounted Cumulative Gain (NDCG@K)~\cite{bao2023bi, rajput2024recommender}, which we compute using the all-ranking protocol~\cite{krichene2020sampled}. 
In this paper, we set K as 5 and 10. 
To better align with real-world scenarios, we follow previous work~\cite{bao2023bi} when dividing the dataset, that is, splitting a dataset into training, validation, and test sets according to timestamps. This ensures that there will be no data leakage issues~\cite{ji2023critical} during the training and testing of the model.

\subsubsection{Baselines}
We adopt the following representative sequential recommendation models as baselines for our experimental comparison:
\begin{itemize}[topsep=0pt,itemsep=0pt,parsep=0pt,leftmargin=*]
    \item[-] Caser~\cite{tang2018personalized} is a method using CNN that models user behaviors through horizontal and vertical convolutional.
    \item[-] GRU4Rec~\cite{hidasi2015session} is an RNN-based method that uses GRU to model the user behavior via encoding the item sequence.
    \item[-] SASRec~\cite{kang2018self} is a method using Transformer to model the item sequences.
    \item[-] GRU4Rec*~\cite{hidasi2015session} is similar to GRU4Rec but uses the LLM-based embedding to initialize the item embedding.
    \item[-] SASRec*~\cite{kang2018self} is similar to SASRec but uses the LLM-based embedding to initialize the item embedding.
    \item[-] TIGER~\cite{rajput2024recommender} is a generative retrieval paradigm for sequential recommendation and introduces a semantic ID to uniquely identify items. We extend to LLMs, using the instruction combined with semantic ID, and generate multiple semantic ID sequences representing items during the recommendation process.
    \item[-] BIGRec~\cite{bao2023bi} is a method that uses instruction-tuning to directly generate items. We have made changes to the decoding method of this approach, ensuring that the produced results are always within the item list of the dataset.
\end{itemize}
\subsubsection{Implementation Details}
For the traditional recommendation models, we optimize them using binary cross-entropy loss and the Adam optimizer with a learning rate searched in $[1e\text{-}2, 1e\text{-}3, 1e\text{-}4]$. We process the data in batches of size 1024, and we adjust the weight decay within the range of $[1e\text{-}2, 1e\text{-}3, 1e\text{-}4, 1e\text{-}5, 1e\text{-}6]$.
For LLM-based methods, for efficiency, we apply Qwen1.5-1.8B~\cite{bai2023qwen} as the backbone LLM. we use the AdamW~\cite{loshchilov2017decoupled} optimizer and adjust the learning rate within the range of $[1e\text{-}3, 1e\text{-}4, 5e\text{-}5]$. During the training, we applied the cosine learning scheduler for 3 epochs and set the early stop patience as one epoch\footnote{In our preliminary experiments, we found that for LLMs trained on the entire dataset, the model generally converges after only 1 to 2 epochs of training.
When we opt to use a language-agnostic model to assist the recommendation model based on LLM in inference, we simply 
 employ the SASRec~\cite{kang2018self}
}.
In our experiments related to the temperature coefficient, we adjusted it within the range of $[1.0, 1.5, 2.0]$. All experiments were conducted on an Ascend 910B with 32GB VRAM.
\input{latex/table/table_ablation}

\subsection{Main Results}

In this subsection, we study the recommendation performance of our \zjz{proposed $D^3$ method} over six open-world datasets.
As shown in Table~\ref{table:main_result}, it showcases our primary experimental results. 
Notably, the results can be categorized into the following groups: traditional models (Caser, GRU4Rec, SASRec), traditional models enhanced by LLM embeddings (GRU4Rec*, SASRec*), TIGER with different decoding methods, recLLM (BIGRec) with different decoding strategies. Regarding recLLM decoding, we investigate not only its default strategy and our $D^3$ approach but also a temperature scaling coefficient strategy (denoted as ``+temp") commonly used to adjust generation scores. As for TIGER, despite not being fully LLM-based, we still apply our method for it to verify our method's wide applicability.
From the table, we have the following observations:


\begin{itemize}[topsep=0pt,itemsep=0pt,parsep=0pt,leftmargin=*]
    \item When compared to the baseline, our decoding approach with recLLM often gets better performance, surpassing all baselines to achieve the best recommendation performance on all datasets. 
    This underscores the significance of studying the decoding strategy for recLLM and demonstrates the superiority of our decoding method which removes the amplification bias and alleviates the homogeneity issue.
    
    
    \item 
    Considering that TIGER also employs beam search for generation during decoding, it may face similar challenges to recLLM. We applied our $D^3$ approach to TIGER. Comparing the results of the original TIGER and the variant with $D^3$ (columns 8 and 9), we observe improvements when using our strategy. 
    This demonstrates the generality of our method, indicating that it is also suitable for non-linguistic, generation-based recommendation systems.

    \item In comparing TIGER (using LLMs but presenting items with semantic IDs) with the recLLM method BIGrec, we find that to leverage LLMs, it is more suitable to directly represent items with text. We conjecture that this may be due to the fact that, while TIGER incorporates information from LLMs during the training of its encoder, using LLMs as encoders can result in the loss of some information. Additionally, using the complete textual representation of items allows for the full utilization of the knowledge learned by the LLM.

    \item We found that simply adjusting the decoding temperature does not improve recommendation performance, although it may enhance recommendation diversity (discussed later in Figure~\ref{fig:diversity}).

\end{itemize}

%% file: latex/table/table1.tex
\begin{table*}[t]
\centering
\caption{
Recommendation accuracy of the compared methods on different-domain datasets.``+Temp'' denotes adjusting the temperature coefficient for BIGRec's decoding. ``+$D^3$'' denotes applying our decoding method to TIGER/BIGRec. The best results are bolded.
}
\resizebox{1.0\textwidth}{!}{
\begin{tabular}{
>{\columncolor[HTML]{FFFFFF}}c |
>{\columncolor[HTML]{FFFFFF}}l |
>{\columncolor[HTML]{FFFFFF}}c 
>{\columncolor[HTML]{FFFFFF}}c 
>{\columncolor[HTML]{FFFFFF}}c |
>{\columncolor[HTML]{FFFFFF}}c 
>{\columncolor[HTML]{FFFFFF}}c |
>{\columncolor[HTML]{FFFFFF}}c 
>{\columncolor[HTML]{EFEFEF}}c |
>{\columncolor[HTML]{FFFFFF}}c 
>{\columncolor[HTML]{FFFFFF}}c 
>{\columncolor[HTML]{EFEFEF}}c }
\toprule
Datasets                                              & Metrics & Caser  & GRU4Rec & SASRec & GRU4Rec* & SASRec* & TIGER  & +$D^3$             & BIGRec & +Temp  & +$D^3$             \\ \midrule
\cellcolor[HTML]{FFFFFF}                              & NDCG@5  & 0.0550 & 0.0562  & 0.0643 & 0.0712   & 0.0672  & 0.0764 & 0.0785          & 0.0813 & 0.0795 & \textbf{0.0816} \\
\cellcolor[HTML]{FFFFFF}                              & HR@5    & 0.0678 & 0.0681  & 0.0715 & 0.0843   & 0.0798  & 0.0853 & 0.0893          & 0.0929 & 0.0897 & \textbf{0.0938} \\
\cellcolor[HTML]{FFFFFF}                              & NDCG@10 & 0.0595 & 0.0614  & 0.0676 & 0.0772   & 0.0731  & 0.0797 & 0.0830          & 0.0856 & 0.0841 & \textbf{0.0871} \\
\multirow{-4}{*}{\cellcolor[HTML]{FFFFFF}Instruments} & HR@10   & 0.0817 & 0.0843  & 0.0817 & 0.1030   & 0.0980  & 0.0958 & 0.1032          & 0.1062 & 0.1040 & \textbf{0.1110} \\ \midrule
\cellcolor[HTML]{FFFFFF}                              & NDCG@5  & 0.0161 & 0.0248  & 0.0477 & 0.0435   & 0.0418  & 0.0484 & 0.0620          & 0.0640 & 0.0513 & \textbf{0.0823} \\
\cellcolor[HTML]{FFFFFF}                              & HR@5    & 0.0224 & 0.0342  & 0.0647 & 0.0566   & 0.0561  & 0.0559 & 0.0748          & 0.0786 & 0.0674 & \textbf{0.1025} \\
\cellcolor[HTML]{FFFFFF}                              & NDCG@10 & 0.0193 & 0.0288  & 0.0535 & 0.0482   & 0.0478  & 0.0512 & 0.0659          & 0.0694 & 0.0579 & \textbf{0.0871} \\
\multirow{-4}{*}{\cellcolor[HTML]{FFFFFF}CDs}         & HR@10   & 0.0485 & 0.0467  & 0.0824 & 0.0715   & 0.0745  & 0.0646 & 0.0868          & 0.0956 & 0.0879 & \textbf{0.1090} \\ \midrule
\cellcolor[HTML]{FFFFFF}                              & NDCG@5  & 0.0122 & 0.0169  & 0.0237 & 0.0282   & 0.0263  & 0.0367 & 0.0394 & 0.0318 & 0.0279 & \textbf{0.0415 }         \\
\cellcolor[HTML]{FFFFFF}                              & HR@5    & 0.0187 & 0.0261  & 0.0338 & 0.0407   & 0.0383  & 0.0495 & 0.0521 & 0.0420 & 0.0379 & \textbf{0.0565 }         \\
\cellcolor[HTML]{FFFFFF}                              & NDCG@10 & 0.0164 & 0.0221  & 0.0290 & 0.0352   & 0.0328  & 0.0417 & 0.0453 & 0.0372 & 0.0355 & \textbf{0.0480}          \\
\multirow{-4}{*}{\cellcolor[HTML]{FFFFFF}Games}       & HR@10   & 0.0321 & 0.0423  & 0.0502 & 0.0624   & 0.0586  & 0.0651 & 0.0706 & 0.0610 & 0.0553 & \textbf{0.0767}          \\ \midrule
\cellcolor[HTML]{FFFFFF}                              & NDCG@5  & 0.0228 & 0.0200  & 0.0356 & 0.0367   & 0.0371  & 0.0421 & 0.0487          & 0.0570 & 0.0497 & \textbf{0.0652} \\
\cellcolor[HTML]{FFFFFF}                              & HR@5    & 0.0316 & 0.0275  & 0.0473 & 0.0518   & 0.0531  & 0.0534 & 0.0610          & 0.0739 & 0.0660 & \textbf{0.0841} \\
\cellcolor[HTML]{FFFFFF}                              & NDCG@10 & 0.0276 & 0.0238  & 0.0398 & 0.0422   & 0.0433  & 0.0465 & 0.0535          & 0.0643 & 0.0565 & \textbf{0.0713} \\
\multirow{-4}{*}{\cellcolor[HTML]{FFFFFF}Toys}        & HR@10   & 0.0465 & 0.0392  & 0.0745 & 0.0688   & 0.0721  & 0.0668 & 0.7600          & 0.0965 & 0.0871 & \textbf{0.1025} \\ \midrule
\cellcolor[HTML]{FFFFFF}                              & NDCG@5  & 0.0439 & 0.0586  & 0.0695 & 0.0577   & 0.0627  & 0.0846 & 0.0876          & 0.0932 & 0.0870 & \textbf{0.0970} \\
\cellcolor[HTML]{FFFFFF}                              & HR@5    & 0.0541 & 0.0663  & 0.0770 & 0.0760   & 0.0785  & 0.0915 & 0.0964          & 0.1040 & 0.0977 & \textbf{0.1083} \\
\cellcolor[HTML]{FFFFFF}                              & NDCG@10 & 0.0469 & 0.0618  & 0.0725 & 0.0624   & 0.0677  & 0.0869 & 0.0901          & 0.0975 & 0.0918 & \textbf{0.1013} \\
\multirow{-4}{*}{\cellcolor[HTML]{FFFFFF}Sports}      & HR@10   & 0.0633 & 0.0761  & 0.0866 & 0.0906   & 0.0939  & 0.0984 & 0.1040          & 0.1171 & 0.1125 & \textbf{0.1215} \\ \midrule
\cellcolor[HTML]{FFFFFF}                              & NDCG@5  & 0.0042 & 0.0060  & 0.0097 & 0.0076   &    0.0103     & 0.0145 & 0.0181          & 0.0182 & 0.0197 & \textbf{0.0217} \\
\cellcolor[HTML]{FFFFFF}                              & HR@5    & 0.0069 & 0.0094  & 0.0146 & 0.0119   &    0.0128     & 0.0183 & 0.0233          & 0.0239 & 0.0253 & \textbf{0.0280} \\
\cellcolor[HTML]{FFFFFF}                              & NDCG@10 & 0.0060 & 0.0078  & 0.0123 & 0.0099   &    0.0151     & 0.0157 & 0.0200          & 0.0204 & 0.0218 & \textbf{0.0240} \\
\multirow{-4}{*}{\cellcolor[HTML]{FFFFFF}Books}       & HR@10   & 0.0123 & 0.0149  & 0.0226 & 0.0189   &    0.0229     & 0.0220 & 0.0290          & 0.0308 & 0.0317 & \textbf{0.0353} \\ \bottomrule
\end{tabular}
}
\label{table:main_result}
\end{table*}


%% file: latex/table/table_ablation.tex
\begin{table*}[ht]
    \centering
        \caption{
    Ablation results for ``$D^3$'' in terms of accuracy, reporting HR@10. ``Baseline'' denotes recLLM's original decoding, serving as a reference here. ``- RLN'' denotes deactivating the operation of $D^3$ on length normalization, i.e., deactivating addressing amplification bias. ``- TFA'' denotes deactivating the use of the text-free model.
    }
    \begin{tabular}{l|cccccc}
    \toprule 
            ~ & Instruments & Books & CDs & Sports & Toys & Games  \\ \midrule
        Baseline & 0.1062 & 0.0308 & 0.0956 & 0.1171 & 0.0965 & 0.0610 \\ 
        $D^3$ & \textbf{0.1111} & \textbf{0.0354} & \textbf{0.1190} & \textbf{0.1215} & \textbf{0.1025} & \textbf{0.0767} \\
        - RLN & 0.1093 & 0.0353 & 0.1000 & 0.1200 & 0.0975 & 0.0659 \\ 
        - TFA & 0.1086 & 0.0309 & 0.1115 & 0.1192 & 0.1006 & 0.0732 \\
         \bottomrule
    \end{tabular}
    \label{table:ablation}
\end{table*}

%% file: latex/6analysis.tex
\section{Analysis}
This section further explores our decoding approach in detail and discusses potential extensions. First, we conduct an ablation analysis to validate the influence of our two key strategies on recommendation accuracy. We then examine the impact of the method's designs on recommendation diversity and investigate an extension of our methods which can enable easy adjustment of the recommendation type distribution. \bk{Finally, we validate the generalizability of the proposed method using additional backbones, approaches, and datasets.}

\subsection{Ablation}

To evaluate the influence of different components of the proposed method \( D^3 \) on accuracy, we conducted ablation studies here. The core elements of \( D^3 \) include 1) removing length normalization to address amplification bias, denoted by ``RLN'', and 2) integrating a text-free assistant model for augmenting diversity, denoted as ``TFA''. We systematically deactivated the components one by one to analyze their effects. The results are summarized in Table~\ref{table:ablation}, where we draw the following conclusions:
\begin{itemize}[topsep=0pt,itemsep=0pt,parsep=0pt,leftmargin=*]

\item[-] Disabling the removal of length normalization (``-RLN'') or the text-free assistant (``-TFA'') model could both lead to decreased performance compared to the complete version of $D^3$. These results emphasize the importance of both designs in addressing their respective issues. Notably, homogeneous recommendations could also limit recommendation accuracy, so the ``-TFA'' model also exhibits decreased performance.

\item[-] When applying ``-RLN'' or ``-TFA'' to $D^3$, only one of the $D^3$ designs is effective, yet it still outperforms the Baseline. This indicates the presence of both the amplification bias and homogeneity issue and demonstrates that addressing them individually can lead to performance improvements. 


\end{itemize}
\input{latex/figure/figure_diversity}
\subsection{Diversity}
As discussed in our previous work, existing generation strategies often result in homogeneous recommendations. In $D^3$, we integrate a text-free assistant to address the issue. Here, we further investigate this design by exploring two research questions:  (1) Can the use of a text-free assistant (TFA) model improve recommendation diversity? (2) How does the effect of TFA combined with diversity-enhancing decoding strategy (temperature adjustment)?
As shown in the Figure ~\ref{fig:diversity}, we have the following findings:
\begin{itemize}[topsep=0pt,itemsep=0pt,parsep=0pt,leftmargin=*]
    \item 
    Adjusting the temperature improves diversity, but as Table~\ref{table:main_result} shows, it results in lower performance.
    \item Using a text-free model for assisted decoding can not only improve the recommendation accuracy but also enhance recommendation diversity.
    \item Utilizing a text-free model for assisted decoding is a plug-and-play approach. It can be combined with temperature adjustment to further improve the diversity of recommendation results.
\end{itemize}

\subsection{Distribution Adjustment}
\input{latex/figure/figure_control}
\input{latex/table/table_generalization}
Taking it one step further, since we can use a text-free model to assist LLMs in decoding, we can also use a similar approach to control the distribution of recommendation results when we have specific recommendation needs. 
\bkq{For instance, we could recommend more computer-related books to users in the computer science field.}
Specifically, this involves strengthening the logits of certain items during the assisted decoding process. 

To validate this extension attribute, we simulated a scenario where we need to directionally enhance the recommendation of a specific group of items and improve recommendation accuracy for that group.
Therefore, when using the text-free model, we only assist items of that particular group. 
As shown in Figure~\ref{fig:control}, we plotted\footnote{Resuls on other three datasets are shown in Figure~\ref{fig:appendix_control}.} the proportion of recommendations for specific categories before and after the control, and the recommendation performance. 
We can see that through this approach, we significantly improved the recommendation proportion and accuracy for specific categories of items. 
This verifies the feasibility of the extension of our method to adjust the recommendation distribution during the decoding stage.

\subsection{Generalizability}
%

To show the generalizability of our methods, we conduct further experiments using additional backbones, approaches, and datasets. In detail, as shown in Table~\ref{table:generalization}, we present the results of applying our decoding method to a new backbone model, Llama3.1-8B~\cite{llama3}, and a new item representation approach, CID~\cite{cid}, on the CDs and Games datasets. 
\input{latex/table/table_steam}
The findings demonstrate consistent improvements in performance with both the new backbone and the new method, highlighting the generalizability and effectiveness of our decoding approach. Besides, we have also applied our approach to the Steam~\cite{kang2018self} dataset. Given the substantial size of this dataset, we randomly selected 10\% of the users to expedite the validation process of our method. The results of our experiments are presented in Table ~\ref{table:steam}. These findings demonstrate that our decoding method is not only applicable to other platforms and datasets but also further validates the generalizability of our approach. 

%% file: latex/figure/figure_diversity.tex
\begin{figure}[t]
    \centering
    \scalebox{0.8}{
    \includegraphics[width=0.55\textwidth]
    {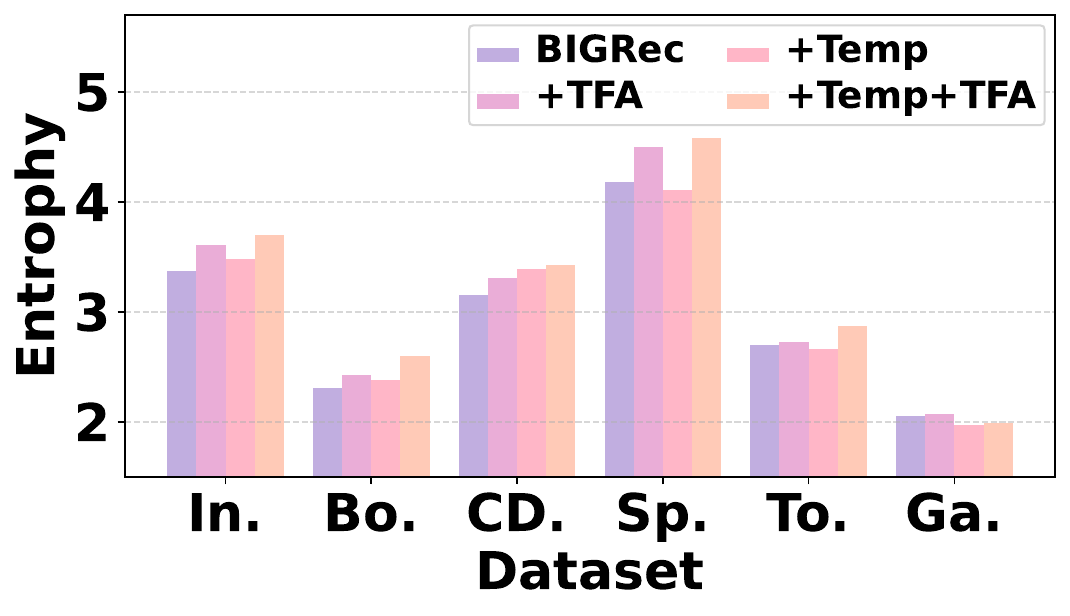}
    }
    \caption{
    Recommendation diversity (measured by entropy) of the original BIGRec and the variants with other decoding strategies. ``+TFA'' denotes the variant applying our text-free model assistant decoding, ``+Temp'' denotes the variant using the widely-used temperature scaling to increase diversity, and ``+Temp+TFA'' denotes the variant combining ``+TFA'' and ``+Temp''. Smaller entropy denotes less diversity.
    }
    \label{fig:diversity}
    \vspace{-0.7em}
\end{figure}

%% file: latex/figure/figure_control.tex
\begin{figure}[t]
    \centering
    \begin{subfigure}[b]{0.235\textwidth}
        \centering
        \includegraphics[width=\textwidth]{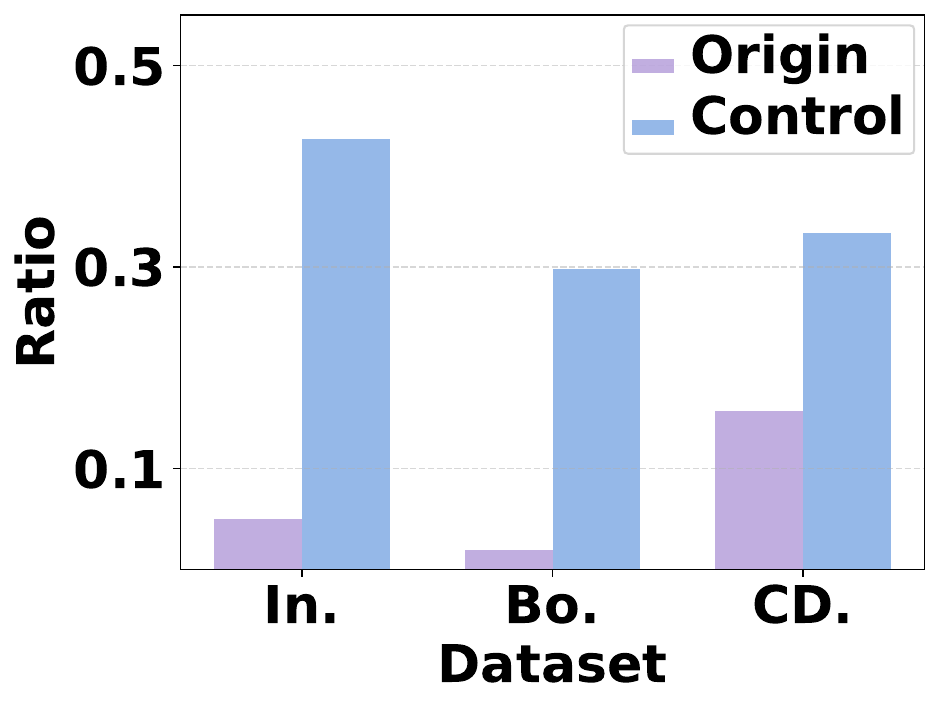}
        \vspace{-1em}
        \label{fig:control_ratio}
    \end{subfigure}
    \begin{subfigure}[b]{0.24\textwidth}
        \centering
        \includegraphics[width=\textwidth]{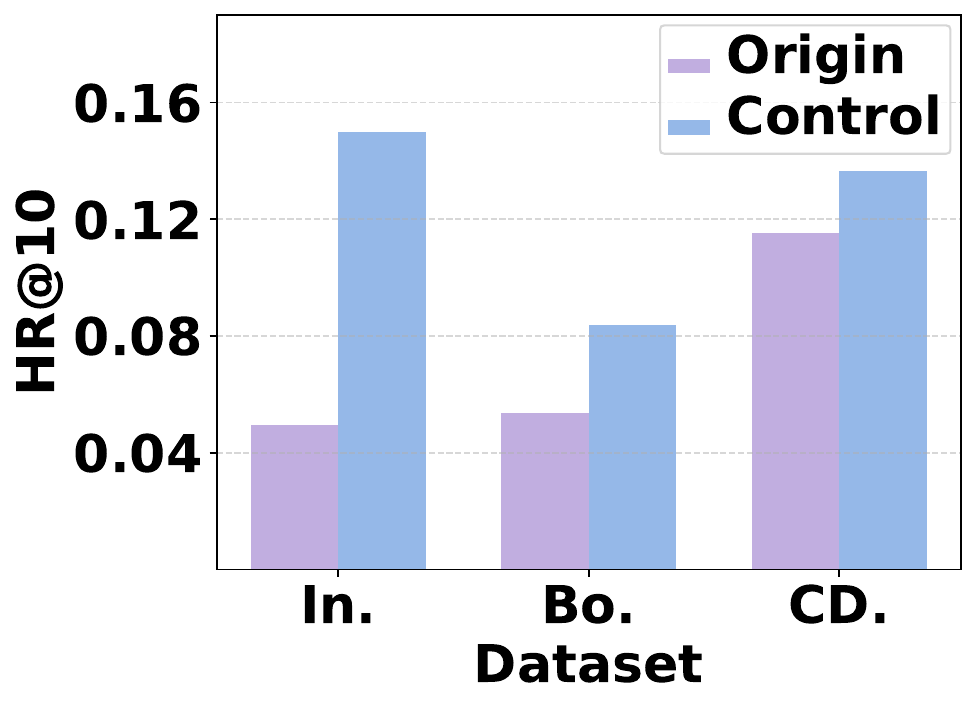}
        \vspace{-1em}
        \label{fig:control_hr}
    \end{subfigure}
    \caption{
    Effectiveness of employing the proposed TFA to enhance recommendation ratio (left) and accuracy (right) for a specified target category of items.
    }
    \vspace{-0.7em}
    \label{fig:control}
\end{figure}

%% file: latex/table/table_generalization.tex
\begin{table}[ht]
    \centering

    \begin{tabular}{>{\columncolor[HTML]{FFFFFF}}l |
>{\columncolor[HTML]{FFFFFF}}c 
>{\columncolor[HTML]{EFEFEF}}c |
>{\columncolor[HTML]{FFFFFF}}c 
>{\columncolor[HTML]{EFEFEF}}c}
    \toprule 
            Datasets & CID & +$D^3$ & BIGRec & +$D^3$ \\ \midrule
        CDs &  0.0610 & \textbf{0.0921} & 0.0869 & \textbf{0.1179} \\ 
        Games & 0.0447 & \textbf{0.0598} & 0.0727 & \textbf{0.0856} \\
         \bottomrule
    \end{tabular}
            \caption{
    We report the HR@10 results of our methods on another approach -- CID~\cite{cid} and another backbone --Llama3.1-8B~\cite{llama3}. ``+$D^3$'' denotes applying our decoding method. The best results are bolded.
    }
    \label{table:generalization}
\end{table}

%% file: latex/table/table_steam.tex
\begin{table}[ht]
    \centering

    \begin{tabular}{>{\columncolor[HTML]{FFFFFF}}l |
>{\columncolor[HTML]{FFFFFF}}c 
>{\columncolor[HTML]{EFEFEF}}c}
    \toprule 
            Metrics &  BIGRec & +$D^3$ \\ \midrule
        NDCG@10 &  0.0277 & \textbf{0.0398} \\ 
        HR@10 & 0.0563 & \textbf{0.0814} \\
         \bottomrule
    \end{tabular}
            \caption{
    We report the NDCG@10 and HR@10 results of our methods on steam datasets~\cite{kang2018self}. ``+$D^3$'' denotes applying our decoding method. The best results are bolded.
    }
    \label{table:steam}
\end{table}

%% file: latex/7conclusion.tex
\section{Conclusion}

In this paper, we begin by conducting a comprehensive analysis of the decoding strategies currently used in LLMs for recommendation, highlighting the critical role of the decoding process. 
During the analysis, we identify two major issues in existing methodologies: 
(1) amplification bias, \textit{i.e.,} amplifying scores for items with ghost tokens,
and (2) homogeneity issue, \textit{i.e.,} 
generating highly similar recommendations and easy-to-produce recommendations with textual features similar to historical interactions 
This work introduces a novel approach, $D^3$, to address the issues by eliminating length normalization and incorporating a text-free model to assist in decoding. Extensive experimental results demonstrate that $D^3$ could significantly enhance recommendation accuracy and diversity. Furthermore, the method can be generalized to non-text generative recommendation frameworks, such as TIGER, indicating its versatility and effectiveness in improving recommendation systems. 

%% file: latex/8appendix.tex

\section{Data Statistic}
\label{appendix:A}
As shown in Table~\ref{table:statstic}, we outline the condition of our dataset. During the preprocessing phase, we took into consideration both the resource implications of training LLMs and the sparsity issues associated with randomly sampled datasets. Consequently, starting from October 2017, we processed the dataset to ensure that each user/item had a minimum of 5-core interactions post-processing. Should the number of items post-processing fall below a threshold (set to 10,000 in our case), we would extend the timeframe by an additional month backward and repeat the procedure, ultimately resulting in six distinct datasets. (Note: Even utilizing the full \textit{Instruments} data, there are only 9,239 items available).
\input{latex/table/table_statistic}
\input{latex/table/table_length_statistic}
\input{latex/figure_app/figure_analyze}
\input{latex/figure_app/figure_control}
\input{latex/figure_app/figure_ratio2}
\section{Analysis of Amplification Bias}
\label{appendix:sec2}

As we have discussed in the main text, due to the specific, non-uniform of the item space in recommendations, there exists a substantial number of \zjz{ghost tokens} during the generation of items. 
These tokens become determinate following the generation of preceding tokens. Therefore, their generation probability is approximately 1, which, under the default decoding strategy, inadvertently inflates the score of the item, thereby introducing bias.
In order to mitigate this issue, we begin by analyzing the length of item representations in terms of tokens within each dataset, as well as the length after removing these tokens, as shown in Table~\ref{table:sss}. Our findings indicate that: 
\begin{itemize}
    \item The proportion of ~\zjz{ghost tokens} significantly impacts the total length, playing a decisive role. 
    \item There is considerable variance in the original token lengths, suggesting that length significantly influences the scores of items during the generation process. 
    \item After the removal of ~\zjz{ghost tokens}, the variance is reduced, and the lengths of items become relatively uniform. Therefore, it becomes feasible to eliminate the length normalization factor directly. 
\end{itemize}

\section{Analysis of Homogeneity Issue}
Figure~\ref{fig:appendix_analysis} illustrates the homogeneity issue observed in the \textit{Sports}, \textit{Toys}, and \textit{Games} datasets, a phenomenon that is consistent with findings in the other three datasets. LLM-based recommender systems demonstrate a significant level of similarity both in terms of text and category. This confirms that across all six datasets, while the integration of textual information enhances performance, it simultaneously introduces a text-level bias. This bias diminishes the diversity of the recommendations, causing them to converge on specific characteristics. 
It's important to clarify that the apparent issue identified within our study is not inherently detrimental. On one side, we note that even text-free models also recommend items with textual similarities, pointing to a fundamental tendency within recommender systems. On the other side, as evidenced by Figure~\ref{fig:app_ratio2}, where we have mapped out the proportion of repetitive item categories within the training set, it's clear that the practice of suggesting items within similar categories inherently fulfills the preferences of certain user demographics. However, when deploying LLM-based recommender systems, their inherent copy mechanisms tend to exaggerate this feature, which might detract from the experience of a broader user base. Consequently, our findings advocate for a strategic adjustment to this issue, aiming for moderation rather than elimination, to balance user satisfaction across the spectrum optimally.
%



\label{appendix:sec3}


%% file: latex/table/table_statistic.tex
\begin{table*}[htbp]

\centering

\begin{tabular}{l|cccccc}
\toprule
      & \multicolumn{1}{l}{Instruments} & \multicolumn{1}{l}{CDs} & \multicolumn{1}{l}{Games} & \multicolumn{1}{l}{Toys} & \multicolumn{1}{l}{Sports} & \multicolumn{1}{l}{Books} \\ \midrule
Item & 9239 & 14239 & 11037 & 11252 & 16003 & 41722 \\
Train & 140482                          & 148685                  & 201613                    & 112755                   & 181477                     & 682998                    \\
Valid & 17561                           & 18586                   & 25202                     & 14095                    & 22685                      & 85376                     \\
Test  & 17562                           & 18587                   & 25203                     & 14096                    & 22686                      & 85376       \\
\bottomrule
\end{tabular}
\caption{The table presents statistical information for six datasets. The first row shows the number of items in our dataset, and the second, third, and fourth rows display the number of sequences in the training, validation, and test sets, respectively.
}
\label{table:statstic}
\end{table*}

%% file: latex/table/table_length_statistic.tex
\begin{table}[htbp]
\centering
\begin{tabular}{ccccc}
\toprule
\multirow{2}{*}{Dataset} & \multicolumn{2}{c}{Origin}                        & \multicolumn{2}{c}{Remove}                        \\ 
\cmidrule(l{3pt}r{3pt}){2-3}
\cmidrule(l{3pt}r{3pt}){4-5}
 & \multicolumn{1}{c}{Avg} & \multicolumn{1}{c}{Var} & \multicolumn{1}{c}{Avg} & \multicolumn{1}{c}{Var} \\ \midrule
Instruments           & 19.34                   & 55.7                     & 4.31                    & 1.67                    \\
CDs                   & 7.64                    & 27.83                    & 3.34                    & 1.15                    \\
Games                 & 12.95                   & 35.16                    & 4.6                     & 2.65                    \\
Toys                  & 15.8                    & 45.41                    & 4.5                     & 2.19                    \\
Sports                & 18.17                   & 68.54                    & 4.12                    & 1.72                    \\
Books                 & 12.27                   & 36.74                    & 3.47                    & 0.84                    \\ \bottomrule
\end{tabular}
\caption{This table presents the average length and its associated variance of the token lists representing items in each dataset before and after removing ghost tokens. }
\label{table:sss}
\end{table}

%% file: latex/figure_app/figure_analyze.tex
\begin{figure*}[htbp]
    \centering
    \begin{subfigure}[b]{0.24\textwidth}
        \centering
        \includegraphics[width=\textwidth]{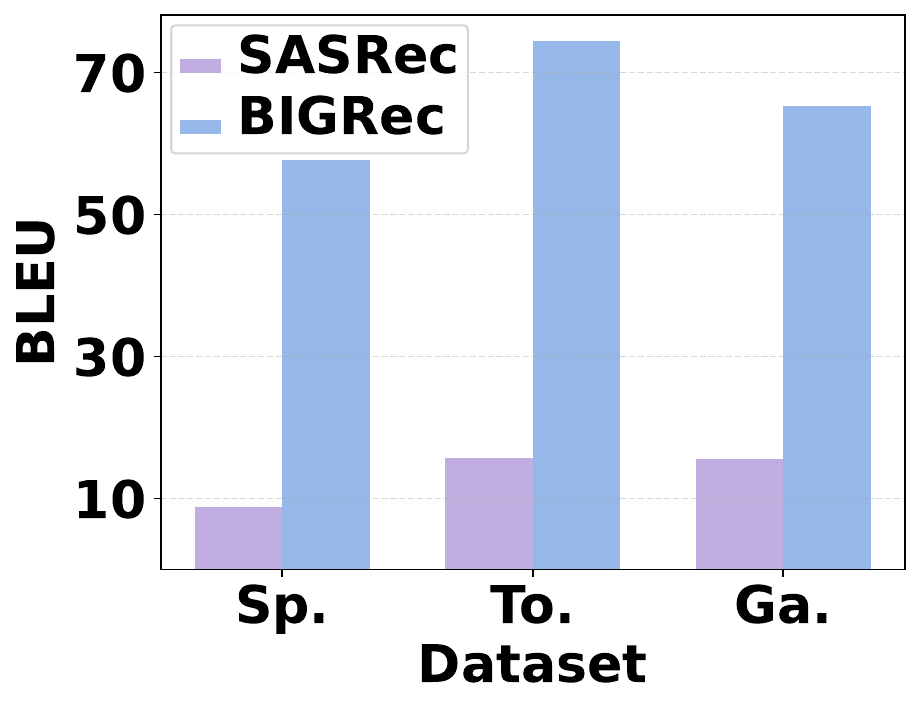}
        \caption{}
        
        \label{fig:app_sub1}
    \end{subfigure}
    \hfill
    \begin{subfigure}[b]{0.24\textwidth}
        \centering
        \includegraphics[width=\textwidth]{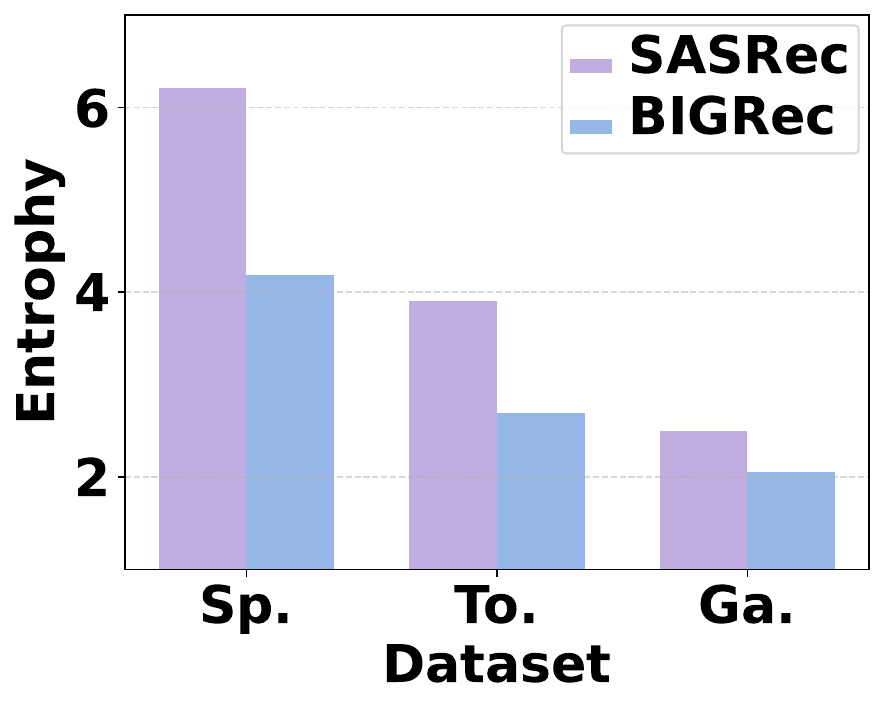}
        \caption{}
        \label{fig:app_sub2}
    \end{subfigure}
    \hfill
    \begin{subfigure}[b]{0.24\textwidth}
        \centering
        \includegraphics[width=\textwidth]{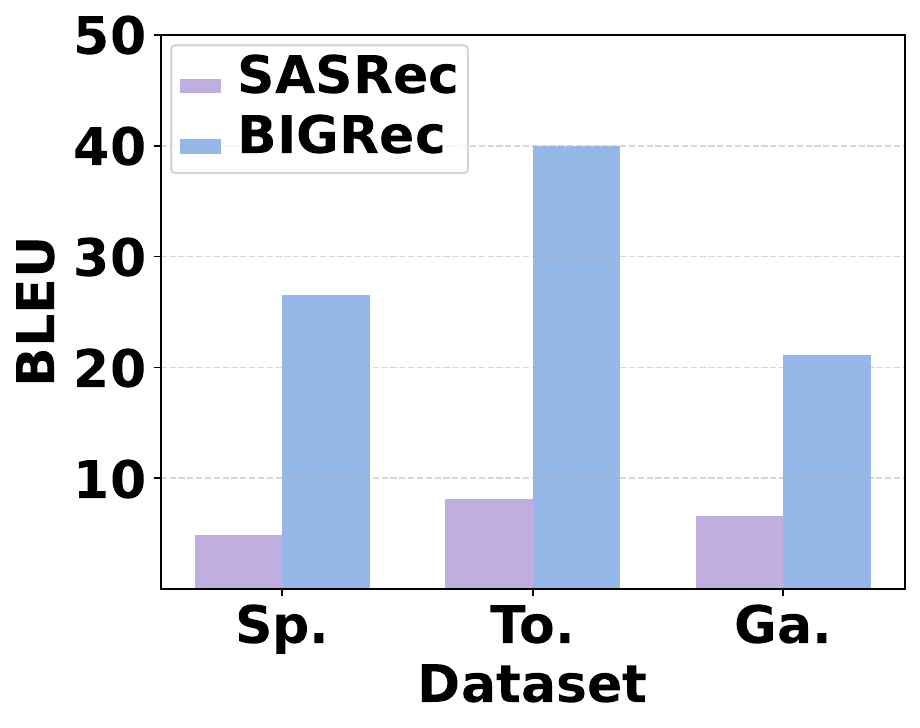}
        \caption{}
        \label{fig:app_sub3}
    \end{subfigure}
    \hfill
    \begin{subfigure}[b]{0.24\textwidth}
        \centering
        \includegraphics[width=\textwidth]{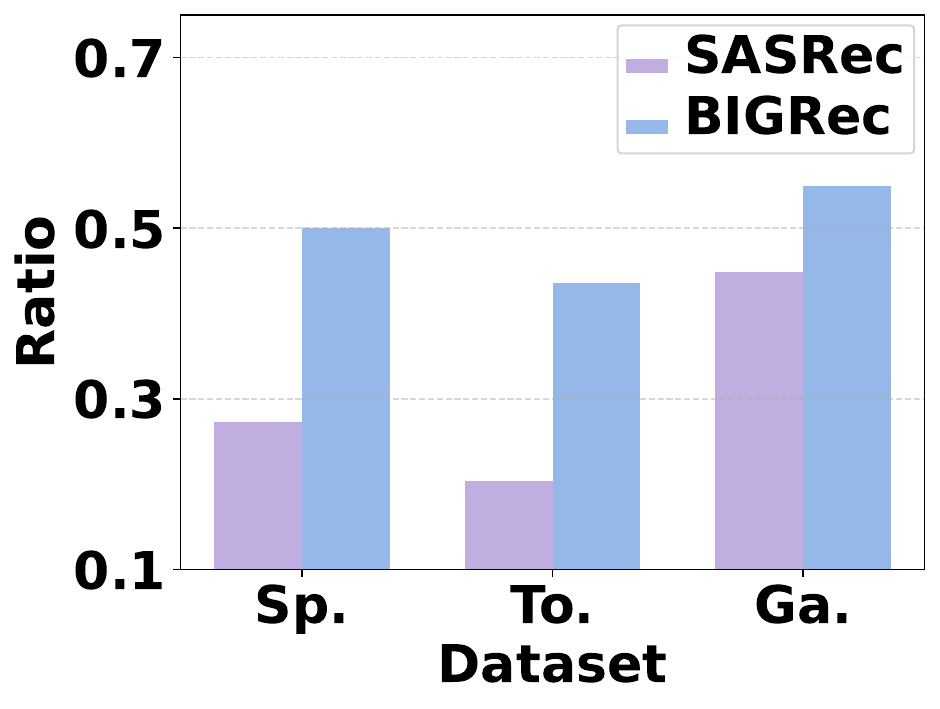}
        \caption{}
        \label{fig:app_sub4}
    \end{subfigure}
    \caption{The analysis of recommendation results using LLMs on the remaining three datasets (abbreviated as \textit{Sp.} for \textit{Sports}, \textit{To.} for \textit{Toys}, and \textit{Ga.} for \textit{Games}) is presented in these four figures. In Figures (a) and (b), the text-similarity of the top 5 tokens within the top 10 recommendations and the entropy of the overall recommendation categories are illustrated, respectively.  Higher text similarity and lower entropy indicate a higher level of homogeneity in recommendations. Figures (c) and (d) depict display text similarity and category repetition in top-10 recommendations versus historical interactions.}
    \label{fig:appendix_analysis}
\end{figure*}

%% file: latex/figure_app/figure_control.tex
\begin{figure}[htbp]
    \centering
    \begin{subfigure}[b]{0.23\textwidth}
        \centering
        \includegraphics[width=\textwidth]{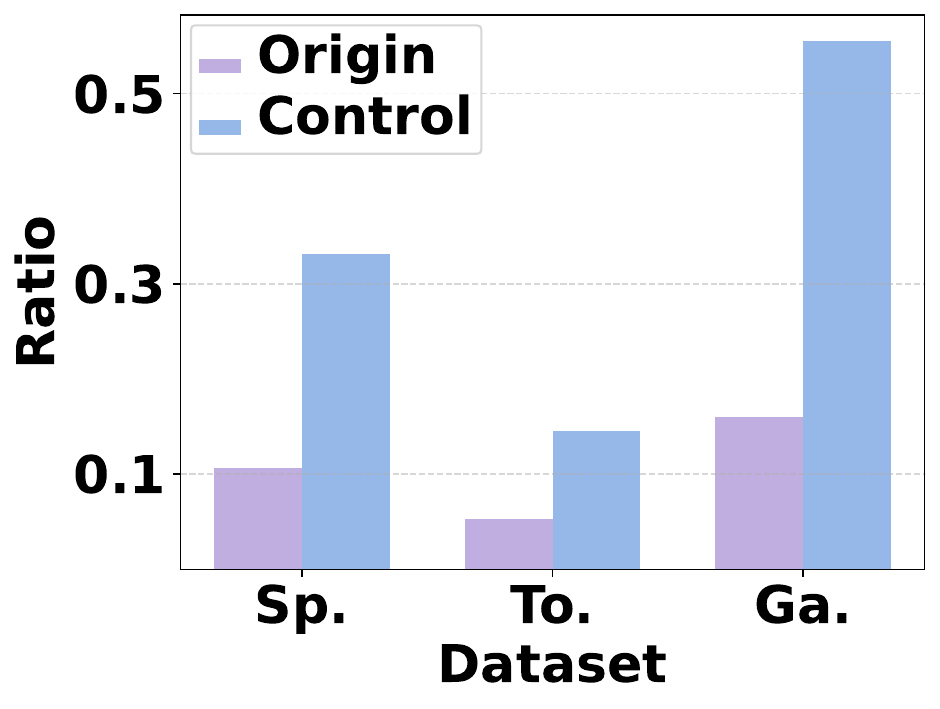}
        \caption{}
        \label{fig:app_control_ratio}
    \end{subfigure}
    \begin{subfigure}[b]{0.23\textwidth}
        \centering
        \includegraphics[width=\textwidth]{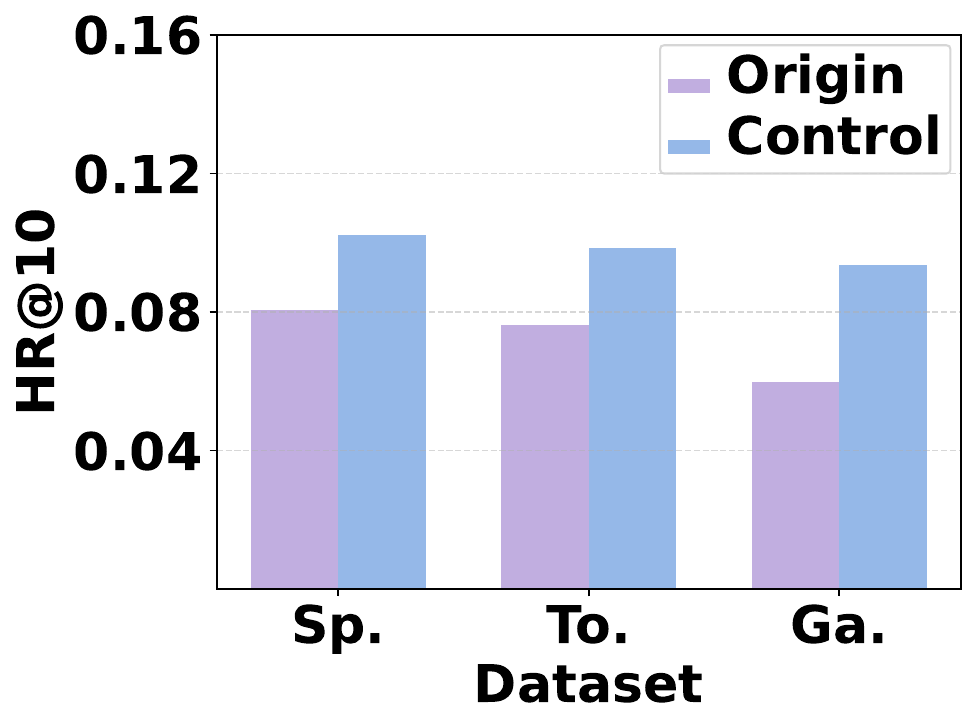}
        \caption{}
        \label{fig:app_control_hr}
    \end{subfigure}
    \caption{These Figures showcase the impact of our proposed TFA method on modifying recommendation distributions. In particular, Figure (a) shows the percentage of recommended items for a particular category after adjustments, whereas Figure (b) depicts the performance of recommendation within that category.}
    \label{fig:appendix_control}
\end{figure}

%% file: latex/figure_app/figure_ratio2.tex
\begin{figure}[ht]
    \centering
    \begin{subfigure}[b]{0.4\textwidth}
        \centering
        \includegraphics[width=\textwidth]{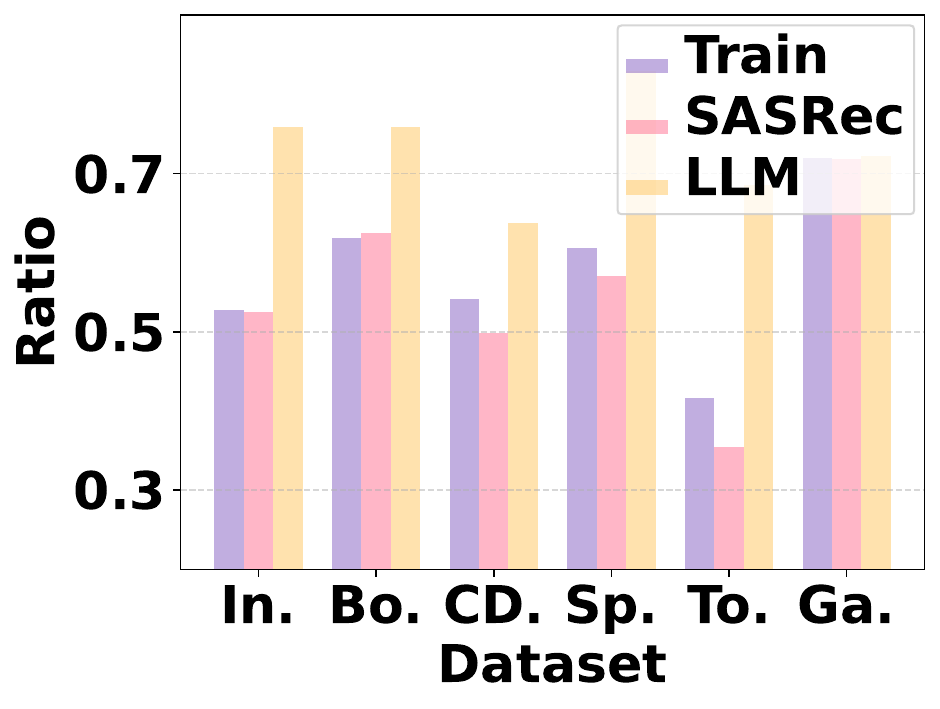}
    \end{subfigure}
    \vspace{-1em}
    \caption{This figure displays the ratio of the training set's ground truth category, item categories of  top-1 recommendations by traditional text-free models and LLM-based recommender systems appears in the user historical interactions. }
    \label{fig:app_ratio2}
\end{figure}